# Electrostatically-induced strain of graphene on GaN nanorods

Jakub Kierdaszuk [1], Rafał Bożek [1], Tomasz Stefaniuk [1], Ewelina Możdzyńska [2,3], Karolina Piętak-Jurczak [2,4], Sebastian Złotnik [2,5], Vitaly Zubialevich [6], Aleksandra Przewłoka [2,7,8], Aleksandra Krajewska [2,7], Wawrzyniec Kaszub [2], Marta Gryglas-Borysiewicz [1], Andrzej Wysmołek [1], Johannes Binder [1], Aneta Drabińska [1]

[1] *Faculty of Physics, University of Warsaw, Poland*
[2] *Łukasiewicz Research Network-Institute of Microelectronics and Photonics, Warsaw, Poland*
[3] *Warsaw University of Technology, Faculty of Materials Science and Engineering, Poland*
[4] *Warsaw University of Technology, Faculty of Chemistry, Poland*
[5] *Institute of Applied Physics, Military University of Technology, Warsaw, Poland*
[6] *University College Cork, College Road, Cork, Ireland*
[7] *CENTERA Laboratories, Institute of High Pressure Physics PAS, Warsaw, Poland*
[8] *Institute of Optoelectronics, Military University of Technology, Warsaw, Poland*

*Corresponding author. E-mail: jakub.kierdaszuk@fuw.edu.pl

Keywords: graphene, gallium nitride, Raman spectroscopy, atomic force microscopy, strain engineering, gating

Abstract

Few-layer graphene deposited on semiconductor nanorods separated by undoped spacers has been studied in perspective for the fabrication of stable nanoresonators. We show that an applied bias between the graphene layer and the nanorod substrate affects the graphene electrode in two ways: 1) by a change of the carrier concentration in graphene and 2) by inducing strain, as demonstrated by the Raman spectroscopy. The capacitance of the investigated structures scales with the area of graphene in contact with the nanorods. Due to the reduced contact surface, the efficiency of graphene gating is one order of magnitude lower than for a comparable structure without nanorods. The shift of graphene Raman modes observed under bias clearly shows the presence of electrostatically-induced strain and only a weak modification of carrier concentration, both independent of number of graphene layers. A higher impact of bias on strain was observed for samples with a larger contact area between the graphene and the nanorods which shows perspective for the construction of sensors and nanoresonator devices.

Introduction

Graphene mechanical nanoresonators and gated structures have been proposed as efficient platforms for the fabrication of optomechanical devices and sensors.[1–4] Recent studies have also underlined their usefulness for mass sensing[5] and tracing physical phenomena in graphene including many-body effects and quasi-ballistic transport.[6–8] In the most widely used capacitor geometry, graphene is deposited on holes etched in a dielectric medium which separates graphene from a conductive substrate.[9,10] Graphene is characterized by a low density of states so an applied external bias can effectively shift the Fermi level.[4,11] Graphene deposited on a semiconductor does not form such a conventional gated structure. However, the notion "gating" is often used in literature to generally describe a structure that allows for the modification of carrier concentration as a function of bias. Moreover, a bias can also alter electrostatic interactions between graphene and the conductive layer affecting the strain in graphene.[9,10] Both phenomena (strain, and change in carrier concentration) can be monitored optically by Raman spectroscopy of graphene.[4,12–14] Bias-controlled graphene strain engineering could also boost the performance of sensors and valley filters based on an interaction between charge carriers in graphene and pseudomagnetic fields.[15–18] Such fields are effectively induced in graphene wrinkles formed in graphene partially supported by nanorods (NRs).[19,20] Graphene transferred on NRs is an interesting candidate structure for the fabrication of sensors with graphene being locally gated by the substrate.[21] The responsivity of such a NR/graphene diode could be one order of magnitude higher than of the planar structure which was interpreted to be caused by a tip-enhanced electric field.[22] Nanorod substrates are also characterized by a lower reflectance, which is promising for the fabrication of light-emitting nano-diodes and solar cells.[23–25] Applying an external voltage to a graphene/NRs junction may both shift the Fermi level of graphene and electrostatically affects the strain distribution.[9] These two parameters should be considered to consistently interpret the properties of such structures, which is important for the construction and optimization of novel devices.

Here we present studies of electrostatically-induced strain and doping in four, three and two-layer turbostratic graphene (4Lgr, 3Lgr, and 2Lgr) on GaN NRs. The turbostratic alignment of the FLgr) used in this study weakens interlayer interactions and coupling, which distinguishes this structure from Bernal stacked graphene.[26–28] At the same time, FLgr electrode offers higher durability and lower resistivity as compared to a graphene monolayer.[24,29] Even if the impact of graphene layer number on its elasticity is weak or negligible, it can affect the charge screening and the possibility of a modulation of the carrier concentration.[30–32] Previous studies of epilayer structure consisting of GaN/4Lgr graphene Schottky diode showed that one can achieve high gating efficiencies (up to $4\cdot10^{11}$ cm$^{-2}$V$^{-1}$ at reverse bias -5 V and $9\cdot10^{12}$ cm$^{-2}$V$^{-1}$ at forward bias 1 V).[11,26] Therefore, an applied electric field at the nanometer scale junction between 4Lgr and GaN is high enough to induce a carrier concentration gradient between the consecutive graphene layers.[26] The combination of the aforementioned phenomena renders a 4Lgr/GaN Schottky diode a promising candidate for applications as sensors and further research including nanorod structures. Therefore, GaN substrate was chosen for the fabrication of lithographically etched nanorods in order to study both effects: i) electrostatically-induced strain and ii) carrier concentration changes via biasing. Furthermore, samples with a different number of graphene layers were investigated to study if this factor affects the strength of interaction with nanorod substrates. The morphology of the obtained structures was analyzed by scanning electron microscopy (SEM) and atomic force microscopy (AFM). Electrical properties were investigated by capacitance-voltage (CV) measurements. Raman spectroscopy was chosen to probe graphene carrier concentration and strain as a function of bias.[4,12,13] Moreover, Kelvin Probe Force Microscopy (KPFM) was applied to study morphology and the distribution of carrier concentration on the surface.[33] We show that the shift of carrier concentration occurs simultaneously with the electrostatically induced strain. In the presented case, strain impacts the energies of Raman modes stronger than doping. Therefore, the aspect of electrostatically induced strain should be considered in the analysis of graphene/NRs devices. This observation also opens pathways for further research in the context of strain engineering and control of pseudomagnetic fields in graphene.

Methods

The GaN heterostructure was grown on a 2-inch single-side polished sapphire ($Al_2O_3$) substrate (maximum off-cut ± 0.25°) by metalorganic vapor phase epitaxy (MOVPE) using an AIXTRON 200/4 RF-S reactor. A 750 nm thick layer of AlN buffer proceeded by a 300 nm thick layer of undoped GaN and a 1.2 μm thick layer of silicon doped n-GaN was grown. Finally, a 100 nm thick layer of GaN was grown on top. The carrier concentration of GaN and n-GaN was approx. $2\cdot10^{16}$ and $4\cdot10^{18}$ cm$^{-3}$, respectively. Then, heterostructures were plasma-etched through a silica sphere lithographic mask to obtain NRs.[34] Silica nanospheres were synthesized by the Stöber method while a selection of synthesis conditions enabled to control of their diameter.[35] Then, one layer of spheres was deposited on the surface as a lithographic mask. Inductively coupled chlorine plasma dry etching was used for NRs fabrications.

Residuals of silica nanospheres were removed in HF solution. As a result, two NRs substrates with an average height of 800 nm were obtained. One was characterized by the average distance between the centers of the nearest NRs of about 230 nm and their average diameter of about 70 nm while the second was characterized by average NRs distance and diameter of 390 nm and 180 nm, respectively. Graphene was grown by chemical vapour deposition (CVD) on copper foil.[36] FLgr was prepared using a several-step procedure. At the beginning (→ as a first step) a graphene monolayer was transferred from a copper substrate onto another graphene monolayer on a copper substrate by a high-speed electrochemical delamination method which forms 2Lgr.[36] Then, a third graphene monolayer was transferred on the previously obtained 2Lgr forming 3Lgr. Finally, a fourth graphene monolayer was transferred on the 3Lgr. The FLgr obtained this way shows a random, turbostratic alignment of graphene layers. Then, a polymer frame method was used to transfer the FLgr from copper onto the NRs substrates.[37] The samples were named by NRs distances and a number (n) of graphene layers (nLNRs230 and nLNRs390). Graphene layers were placed in the centres of the samples to avoid short circuits. Two indium ohmic contacts to the bottom n-GaN layer were made on each sample while silver paint was used to contact graphene. At forward bias, a higher potential was applied to graphene. Schematic drawings/descriptions of samples and contact positions are presented in Fig. 1a, b.

CV measurements were performed using Agilent E4980A impedance meter. The AC amplitude was 10 mV while frequencies were different for each sample.. Explanation of chosen frequencies and results of impedance magnitude and phase measurements as a function of frequency are discussed in the Supplementary Materials. Raman spectra were measured with a Renishaw InVia spectrometer equipped with a 532 nm laser excitation source and 100× objective. The excitation power was reduced to a few mW to minimize heating effects. The laser spot diameter was around 500 nm. Therefore, even in the nLNRs390 sample, the laser spot covers both graphene on NRs and suspended graphene areas. All AFM measurements, topography, and surface potential, were performed using a Bruker Dimension Icon AFM with a Nanoscope VI controller. The probes used for KPFM scanning were FMG01/Pt (NT-MDT) with a tip radius of less than 35 nm. During KPFM measurements, a bias was applied to the sample from an electric battery connected to a potentiometer and multimeter.

Results

Figures 1c-f show the morphology of 4LNRs230 and 4LNRs390 samples which are similar to 3L and 2L samples (Fig. 1a, b). NRs in both samples are organized in a quasi-hexagonal lattice

but a more regular arrangement characterizes the NRs390 substrate (Fig. 1d, f). The surface contrast on NRs390 facets suggests that their top surface is rougher than for NRs230 top facets (Fig. 1c, e). AFM images (Fig. 1e, f) of NRs are consistent with SEM results (Fig. 1c, d). The analysis of an AFM profile from sample 4LNRs230 substrate (Fig. 1e, g) shows that NRs for this sample are rounded while 5-10 nm deep recesses are present on the top of each NR in the 4LNRs390 sample (Fig. 1f, h). NR dimensions obtained from the analysis of SEM images are presented in Fig. 1a, b. Comparison of AFM profiles measured on bare NR substrates and 4LNRs samples confirms that samples differ in the percentage of graphene supported by NRs (more details are presented in Supplementary Materials). We estimate that 13% and 29% area of graphene in contact (AGC) with GaN NRs in 4LNRs230 and 4LNRs390 samples, respectively (Tab. 1).

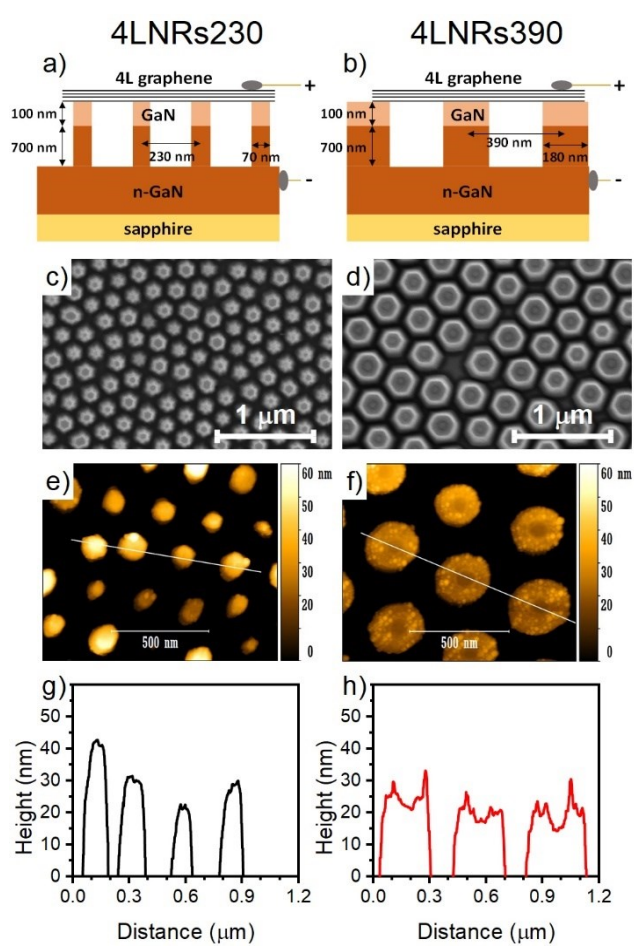

Figure 1. Schemes presenting samples 4LNRs230 and 4LNRs390: a) and b), respectively. + and – show potentials for forward bias. Topography of the 4LNRs230 sample: c) SEM image of NRs (top view), e) AFM image, g) AFM profile from e); Topography of the 4LNRs390 sample: d) SEM image of NRs, f) AFM image, h) AFM profile from f).

The capacitance dependences on bias between the top and bottom electrodes (see Fig.1 a)) are presented in Figure 2. The value of the capacitance of the FLgr/NRs junction in the NRs390 sample measured at 0 V bias is about 1.8 times larger than in NRs230 (Tab. 1).

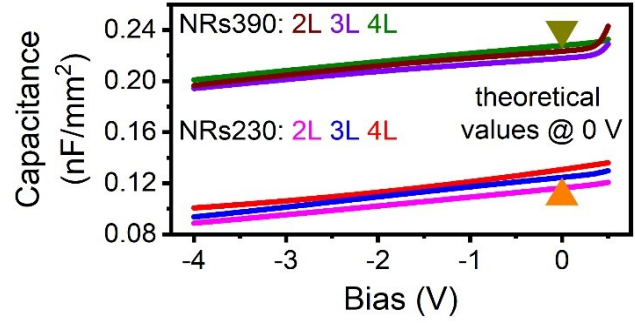

Figure 2. CV characteristics of samples nLNRs230 and nLNRs390 with different number of graphene layers. Triangles indicate the model values of capacitance at 0 V for FLgr at NRs 230 and 390, respectively.

Table 1. Comparison of the area of graphene in contact with NRs obtained from AFM (AGC), surfaces area of graphene electrodes, model values of capacitances at 0 V ($C_{mod.}$), and experimental capacitances at 0 V ($C_{exp}$). $C_{mod.}$ was calculated from Equation (1), see below.

| Sample | AGC (%) | $S$ (mm$^2$) | $C_{mod.}$ (nF/mm$^2$) | $C_{exp.}$ (nF/mm$^2$) |
|---|---|---|---|---|
| 2LNRs230 |  | 14.5 |  | 0.12 |
| 3LNRs230 | 13 | 20.6 | 0.11 | 0.12 |
| 4LNRs230 |  | 13.7 |  | 0.13 |
| 2LNRs390 |  | 18.1 |  | 0.22 |
| 3LNRs390 | 29 | 17.6 | 0.24 | 0.22 |
| 4LNRs390 |  | 18.1 |  | 0.21 |

A quasi-linear dependence of capacitance on bias is present for all samples. The investigated structures at zero bias could work as a capacitor (Fig. S1), while the difference in the area of graphene in contact with NRs affects the value of the capacitance. We tested this hypothesis with a simple model of a parallel plate capacitor. In this concept, we assume that 4Lgr and n-GaN conductors are separated by a $d_1$ = 100 nm GaN layer with $p$ as the effective electrode area, which is the NRs/graphene contact area percentage concerning the whole electrode area

*S*. (1- *p)* percent of conductor area is separated by *d₂* = 800 nm thick air separator. The capacitance of the whole junction can be approximated as:

$$C_{mod} = \frac{\varepsilon_0 \varepsilon_{rGaN} S p}{d_1} + \frac{\varepsilon_0 \varepsilon_{rair} S(1-p)}{d_2} \qquad (1)$$

and depends on the vacuum permittivity ($\varepsilon_0$ = 8.85·10$^{-12}$ F·m$^{-1}$) and the static permittivity of GaN ($\varepsilon_{rGaN}$ = 8.9) and air ($\varepsilon_{rair}$ = 1).[38–40] The results are presented in Table 1. The values of capacitance per mm$^2$ obtained from CV measurements are in good agreement with the model predictions. Thus, FLgr on GaN/n-GaN NRs behaves like a capacitor which is consistent with the previous results regarding 4Lgr deposited on a GaN/n-GaN epilayer heterostructure.[26] Small changes of capacitance as the function of bias could be related to the Schottky-Mott relationship.[11,41]

The impact of the area of graphene in contact with NRs on the evolution of Raman spectra of 4Lgr are analyzed in the next three Figures 3-5. Figure 3 shows a comparison of Raman spectra for samples 4LNRs230 and 4LNRs390. Raman spectra have similar energies of G and 2D bands at zero bias as well as ratios of band intensities (Fig. 3a). Both G and 2D bands consist of one band component. Further analysis showed that samples can differ in the number of G and 2D band components and subbands, which could be an effect of carrier concentration gradient at subsequent graphene layers.[26] This topic is discussed in detail in the Supplementary Materials (Fig. S3, S4 and S5). Interestingly, the average value of the G band at 0 V depends on the number of graphene layers (Fig. S6a). Similarly to Fig. 4, the highest value is characteristic of 2Lgr on NRs while the lowest is present in 4L gr on NRs. This effect could be related to the less effective charge transfer between NRs and top graphene layers caused by the screening effect in thicker graphene electrodes. However, all G and 2D bands in both samples show a blueshift with increasing applied reverse bias. In the range of 1 V to -3V, the shift for the 4L_NRs390 sample is comparable to 4L_NRs230, but at higher biases, the shift becomes larger. Similar trends are present in 3L and 2L samples. The statistical analysis showed a small effect of the number of graphene layers on the average values of G and 2D band shifts which is discussed in supplementary materials (Fig. S6).

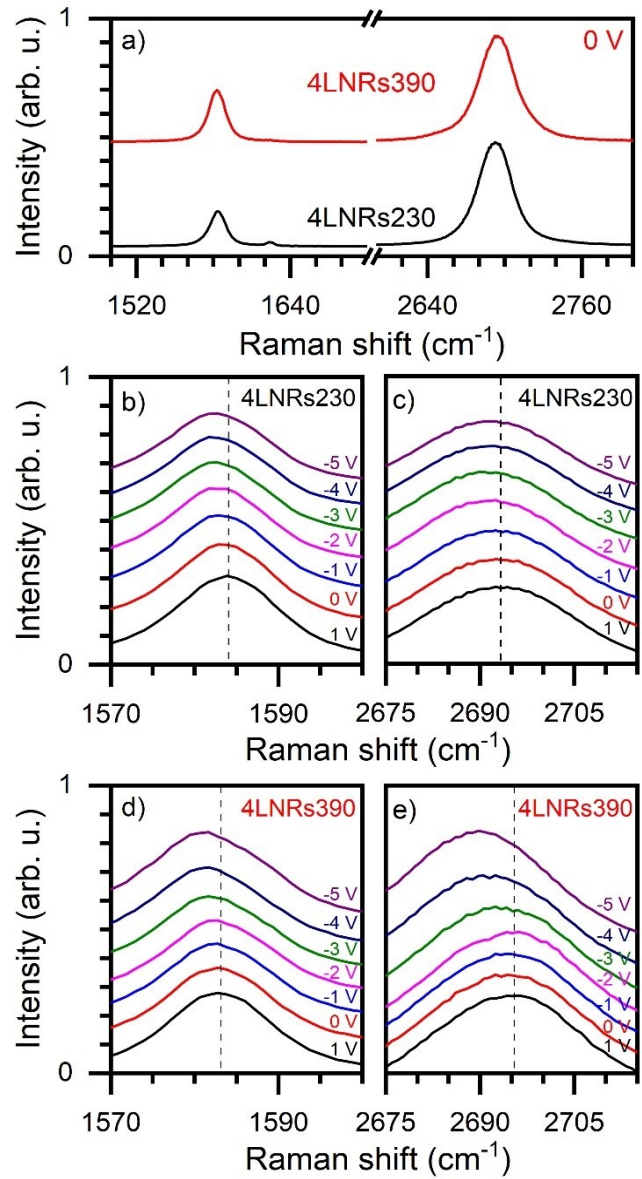

Figure 3. a) representative Raman spectra for samples of 4L at NRs measured at 0 V; b) G and c) 2D band evolution for sample 4LNRs230; d) G and e) 2D band evolution for sample 4LNRs390. Vertical dashed lines indicate G and 2D band energies at 1 V.

Previous research showed that a charge transfer at the Schottky diode junction induces n-type doping in graphene deposited on n-type GaN.[11,42] However, interaction with adsorbed gases in the suspended graphene could also induce p-type doping of graphene.[43,44] The predominance of suspended graphene areas for samples 4LNRs230 and 4LNRs390 could be responsible for p-type doping. In this doping regime, a reverse bias should decrease the p-type doping, which should be correlated with a simultaneous decrease of 2D and G band energies.[4,11,45] However, these parameters also depend on graphene strain. The influence of both, doping and strain, can be distinguished by an analysis of the slope of the dependence of the G band energy on the 2D band energy $E_G(E_{2D})$.[12,13] The slope of the $E_G(E_{2D})$ dependence of around 0.45 is characteristic of biaxial strain, while a more significant slope (around 1.3) characterizes p-type doping.[13,46] These coefficients are characteristic of graphene monolayer, however, they should also apply to the description of turbostratic graphene, where the interlayer coupling is weak.

Figure 4 presents $E_G(E_{2D})$ dependence for a representative series of spectra measured for all samples. Based on the G and 2D band energies we concluded that graphene on NRs is compressively strained. This type of strain was previously explained by taking into account the morphology of graphene on NRs[20]: the laser spot covers an area of compressively strained graphene stretched over NRs, tensely strained graphene wrinkles, and relaxed suspended graphene located between NRs and wrinkles. Moreover, because of Poisson's effect, tensile strained wrinkles are surrounded by compressively strained graphene. Consideration of all the mentioned factors made it possible to explain the occurrence of compressive strain.

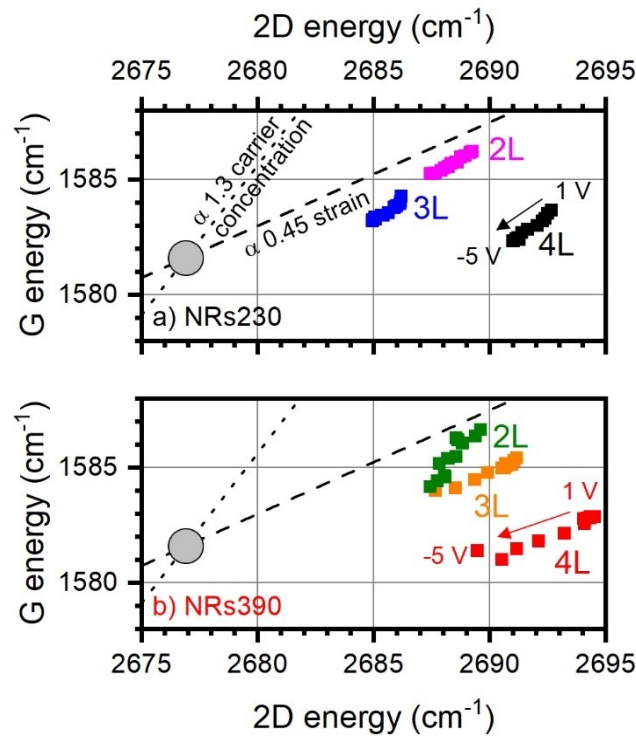

Figure 4. The dependence of G band energy on 2D band energy for representative spectra measured for 2L, 3L, and 4L graphene on nanorods: a) NRs230 and b) NRs390. The dotted line describes the trend of a change in carrier concentration, while the dashed line describes the trend of biaxial strain.[13] The grey dot indicates G and 2D band energies for unstrained and undoped graphene.

Moreover, an increase of reverse bias results in decreasing of both band energies with the trend line characteristic for biaxial strain. Therefore, increasing the reverse bias leads to the generation of tensile strain, which decreases the absolute value of strain in graphene. This effect occurs even if 2-3 G or 2D band components are present (Fig. S3). Furthermore, the average value of 2D band shift for nLNRs230 and nLNRs390 is equal to 1.86 and 3.23 cm$^{-1}$, respectively. These shifts allow to estimate the strain reduction in the NRs230 samples by a factor of 0.03% and 0.05% in the NRs390 samples.[12] The ratio of strain changes in NRs230 and NRs390 samples induced by a reverse bias is 0.48, similar to the ratio of AGC (area of graphene in contact with GaN NRs) equal to 0.45. This finding suggests that the average value of electrostatically induced strain in graphene on NRs depends on the area of graphene in contact with NRs.

Studies of the G band full width at half maximum ($FWHM_G$) and the intensity ratio of 2D and G band ($R_{2DG}$) were performed to check if the bias-induced strain occurs together with a change of carrier concentration (Fig. 5). For smaller carrier concentrations the phonon lifetime decreases and an increase of $FWHM_G$ is observed.[4] The 2D band intensity is reduced by scattering on free carriers, while the G band intensity is independent of the carrier concentration.[47] Therefore, both the $FWHM_G$ and the $R_{2DG}$, inversely depend on the graphene carrier concentration.[4,48] $FWHM_G$ values at 0 V for nLNRs230 are in the range of 15-16 cm$^{-1}$, which is characteristic of a low doping level.[4] Slightly lower values of $FWHM_G$ 12-15 cm$^{-1}$ characterize nLNRs390. Reverse bias increases the $FWHM_G$ of sample nL_NRs390 (Fig. 5b) by a higher value as compared to nL_NRs230 (Fig. 5a). A slight increase of $R_{2DG}$ value at reverse bias is also observed. However, changes in this parameter are comparable for both types of NRs (Fig. 5c, d). Thus, graphene on GaN NRs is initially slightly p-type doped, and reverse bias decreases p-type doping. The value of carrier concentration modification for the bias range between 1 V and –5 V is of the order of $10^{11}$ cm$^{-2}$ for both samples[9], although a slightly larger modification of carrier concentration occurs for sample nLNRs390 characterized by a larger area of graphene in contact with NRs. The efficiency of gating in graphene on GaN/n-GaN NRs structure is at least one order of magnitude lower than on GaN/n-GaN epilayer.[26] Therefore, electrostatically induced strain in the nLNRs sample is dominant, however, a small gating effect is also present.

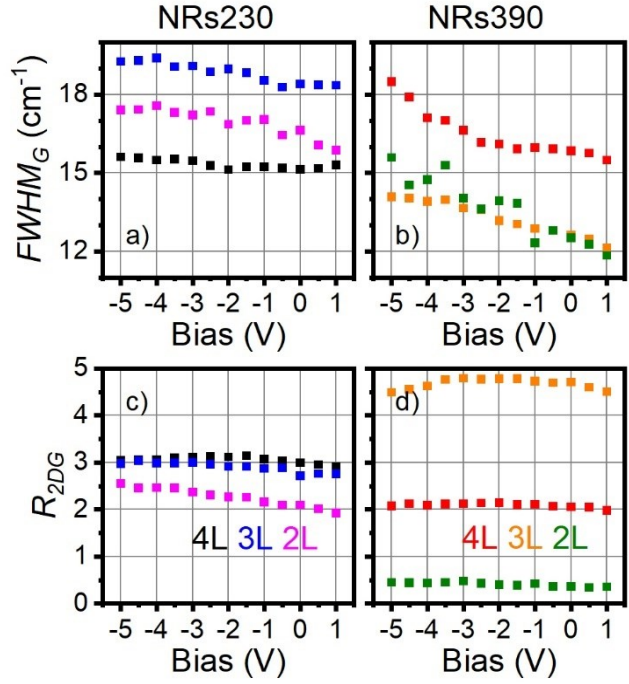

Figure 5. $FWHM_G$ dependence on bias in samples: a) 4L_NRs230, b) 4L_NRs390, the intensity ratio of 2D and G band $R_{2DG}$ in samples: c) 4L_NRs230, d) 4L_NRs390.

The effect of graphene gating was also studied directly by KPFM. These measurements were performed twice for a wide range of biases in the sequence: +1 V, 0.05 V, –1 V, –3 V, –5 V, 1 V, 0.05 V, –1 V, –3 V. Figure 6 shows results for samples of 4Lgr on NRs performed at two different biases 50 mV and –1 V. Images at higher reverse biases were difficult to analyze due to the increase of electrostatic repulsion force between graphene and KPFM tip. The topography of 4Lgr in both samples at bias 50 mV looks similar to results obtained at –1 V if one accounts for a drift (Fig. 6a-d). Graphene wrinkles connect the neighbouring GaN NRs. The KPFM topography of graphene wrinkles for the 4L_NRs230 sample has lower regularity (Fig. 6a, b), which is related to a less uniform distribution of GaN NRs on the surface (see SEM images in Fig. 1c, d). In some samples, a local correlation with the presence of GaN nanorods was observed (Fig. 6g, h). No correlation with the position of supporting GaN NRs in the 4LNRs230 sample is present even in the second series of scans. Measurements performed in other samples of 3Lgr and 2Lgr on GaN NRs showed that local modulation of carrier concentration is present, however, their distinctness varies between samples (Fig. S7). No implicit correlation between the surface potential and the number of graphene layers was observed (Table 2). The values of surface potential for nLNRs390 are slightly higher than for nLNRs230.

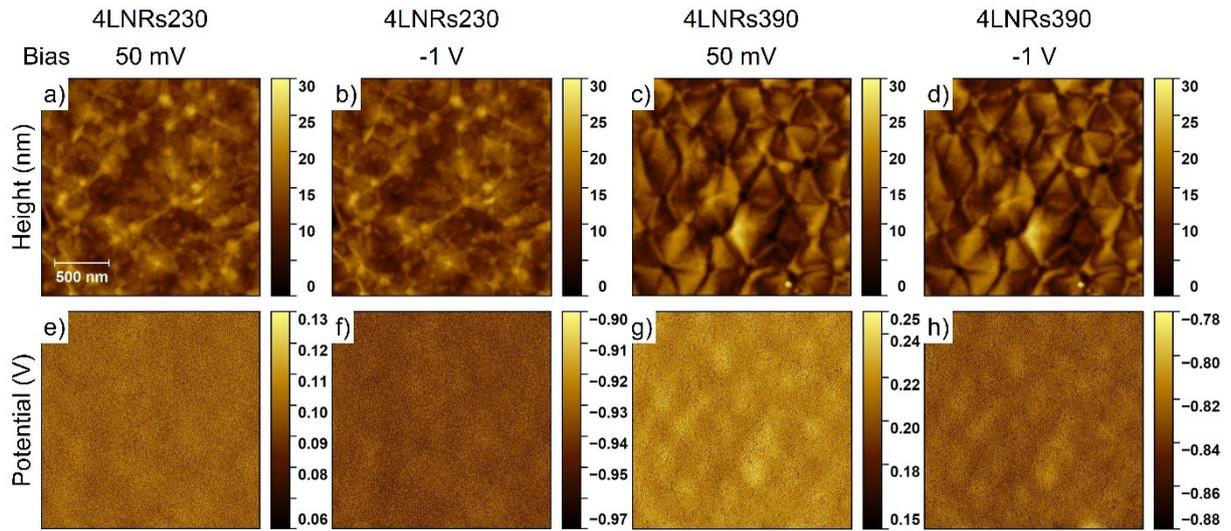

Figure 6. KPFM results obtained during the second scan including sample height: a) 4LNRs230 at 50 mV, b) 4LNRs230 at −1 V, c) 4LNRs390 at 50 mV, d) 4LNRs390 at −1 V; KPFM results of potential distribution on the graphene surface: e) 4LNRs230 at 50 mV, f) 4LNRs230 at −1 V, g) 4LNRs390 at 50 mV, h) 4LNRs390 at −1 V. XY scale in all images is the same.

Table 2. The average value of potential ($V_{av}$) measured at biases 50 mV and −1 V for 4Lgr, 3Lgr, and 2Lgr_on NRs230 NRs390 substrates:

|  | Bias 0.05 V | Bias −1 V |
|---|---|---|
| Samples | $V_{av}$ (mV) | $V_{av}$ (mV) |
| 4LNRs230 | 49 | 62 |
| 3LNRs230 | 164 | 161 |
| 2LNRs230 | 146 | 152 |
| 4LNRs390 | 166 | 171 |
| 3LNRs390 | 190 | 192 |
| 2LNRs390 | 190 | 196 |

The above observation suggests that charge screening could be responsible for a lack of dependence on this value on the number of graphene layers.[32] Furthermore, the electrostatically induced strain could affect mainly the bottom graphene layer. Sliding of that layer under the top ones explained the independence of the KPFM morphology on the bias. Therefore, measurement of a large area of a graphene monolayer on NRs is necessary for better imaging investigated effects. This will precisely show if electrostatically induced strain effect graphene morphology and local distribution of carrier concentration in a detectable range.

However, obtaining a large area of a graphene sheet on rough nanorods without cracks as well as the fabrication of electric contact for fragile graphene monolayer will be challenging. Electrostatically induced graphene strain in 4Lgr, 3Lgr, and 2Lgr was too small to be convincingly observed by AFM even if this phenomenon was easily recognizable by Raman spectroscopy. The mechanical stability of FLgr was confirmed as promising for further optimization for sensing applications.

Conclusions

A combination of several experimental techniques shows the occurrence of electrostatically induced strain and gating in four, three and two-layer turbostratic graphene on GaN NRs. Electrical transport measurements show that the value of the sample capacitance depends on the percentage of graphene area supported by NRs. Raman spectra measured as a function of bias show the simultaneous shifts of G and 2D band energies which is interpreted in terms of an electrostatically induced strain effect. Changes in the carrier concentration of graphene on GaN NRs are one order of magnitude weaker than for a flat epitaxial GaN layer. Raman results show that the gating phenomenon affects Raman band energies of partially suspended graphene less than the strain present in the structure. No explicit correlation was found between the strain value and the number of graphene layers. The electrostatically induced graphene strain value depends on the contact area between graphene and NRs. KPFM measurements also showed the presence of a small carrier concentration shift. However, the screening effect is probably responsible for the independence of carrier concentration distribution on the number of graphene layers. Thus, our studies showed that for understanding gated structures composed of partially suspended graphene and for its application both effects, electrostatically induced strain and gating, should be considered.

Acknowledgements:

The research leading to these results has received funding from the Research Foundation Flanders (FWO) under Grant no. EOS 30467715.

Karolina Piętak-Jurczak acknowledges financial support from the IDUB project (Scholarship Plus programme).